\documentclass{ws}

\def \pl {\partial}
\def \lf {\left (}
\def \rt {\right )}

\def \beq {\begin{equation}}
\def \eeq {\end{equation}}
\def \To {\Rightarrow}

\begin{document}

\begin{center}

{\hbox to \hsize{\hfill CU-TP-1105}}

\bigskip

{\Large \bf Energy Conservation and Hawking Radiation}

\bigskip
\bigskip
\bigskip

{\large \sc Maulik K. Parikh{\small} 
\footnote{e-mail: {\tt mkp@phys.columbia.edu}}\\[1mm]}

{\em Department of Physics, Columbia University, New York, NY 10027}\\[3mm]

\vspace*{1.5cm}

{\large \bf Abstract}
\end{center}
{\large The conservation of energy implies that an isolated radiating black
hole cannot have an emission spectrum that is precisely thermal. Moreover,
the no-hair theorem is only approximately applicable. We consider
the implications for the black hole information puzzle.
}\footnote{Based on a talk presented at the Xth Marcel Grossmann
  Meeting, held in Rio de Janeiro, July 20-26, 2003. To appear
  in the proceedings.}
\noindent

\section{Introduction}

Stephen Hawking's astounding discovery\cite{hawking} that black holes
radiate thermally set up a disturbing and difficult problem: what
happens to information during black hole evaporation? Taken literally,
Hawking's result implies the loss of unitarity or, to put it more
dramatically, the breakdown of quantum
mechanics\cite{infoloss}. Although derivations in string theory
support the idea that Hawking radiation can be described within a
manifestly unitary theory,\cite{callanmalda} it remains a mystery how
information is returned. One suggestion is that
nonlocality should play an important role; indeed, one expects on
general grounds that canonical commutation relations should be
modified in the presence of gravity\cite{kvv}. Here we will explore
an alternative hope: that perhaps energy conservation or, more
generally, gravitational back-reaction provides a loophole.

At a macroscopic level, the claim of information loss in black hole
radiance rests on two pillars: numerous derivations showing that black
holes have an exactly thermal emission spectrum, and the validity of the
no-hair theorem. A thermal spectrum is entirely determined by
specifying a single number, the temperature. So, if exact thermality
holds, the outgoing radiation does not have any information.
Meanwhile, the no-hair theorem asserts that the geometry
outside a stationary black hole is entirely specified by a small
handful of parameters: the mass, the charge, the angular momentum, and
any other Noether charges. (Indeed, these charges also determine the
Hawking temperature). So the spacetime geometry doesn't carry
much information either. But if neither the geometry nor the radiation
carries any information, then, once the black hole has evaporated,
there are no macroscopic signatures of the collapsed matter left.
(Of course, a loss of macroscopic information is not in
conflict with quantum mechanics. But if we could show that there are
coarse-grained features of the outgoing radiation
that correlate with the configuration of the collapsed matter, it
would demonstrate that at least some information is returned without
having to solve the full quantum gravity problem.)

Upon a moment's reflection, however, we see that both thermality and
hairlessness cannot be taken at face value. Either condition,
if strictly true, would violate the conservation of energy. A thermal
spectrum contains a tail of arbitrarily high energies, but 
an isolated black hole obviously cannot emit a particle with more energy than
the mass of the hole. So energy conservation guarantees that as one
goes to higher energies, the spectrum must start deviating from thermality. Put
another way, in a microcanonical ensemble, temperature is only a
low-energy approximation. Moreover, the conservation of energy demands
that, as a quantum of radiation is emitted, the left-over mass of the
black hole must decrease, and the hole must shrink. So the no-hair
theorem, which describes the possible configurations of unchanging
stationary black holes, applies only approximately.

Indeed, energy conservation is not merely a technical detail that
needs to be respected. Rather it is what drives the dynamics:
a black hole radiates because it can
lower its mass. This supports the idea that, in quantum gravity, one
should regard a nonextremal black hole as an excited, metastable
state. We therefore need a derivation of Hawking radiation that is suited
to enforcing energy conservation i.e. for which the spacetime geometry
is dynamical. One such derivation\cite{tunnel}
directly implements
Hawking's heuristic picture of the radiation as particles tunneling
across the horizon. (An alternate viewpoint, in which the radiation is
regarded as the spontaneous emissions of a
membrane\cite{membrane} living at the horizon is also
possible.) Here we will show that the radiation can indeed be viewed
as tunneling particles and that this leads to nonthermality. The
corrected emission rate may plausibly lead to short-time correlations
in the spectrum.

\section{Painlev\'e Coordinates}

To describe tunneling we need a coordinate system that, unlike Schwarzschild
coordinates, is regular at the horizon; particularly convenient are
Painlev\'e coordinates. Consider then a general static metric of the form
\beq
ds^2 = - (1 -g(r)) dt_s^2 + {dr^2 \over 1 - g(r)} + r^2 d
\Omega_{D-2}^2 \; .
\eeq
For a Schwarzschild black hole in four dimensions, $g(r) = 1 - 2M/r$.

To obtain the new line element, define a new time coordinate, $t$, by
$t_s = t + f(r)$. The function $f$ is required to depend only on $r$
and not $t$, so that the metric remains stationary, i.e.
time-translation invariant. Stationarity of the metric automatically
implements the desirable property that the time direction be a Killing
vector. Now, our key requirement is that the metric be regular at the
horizon. We can implement this as follows. We know that a radially
free-falling observer who falls through the horizon does not detect
anything abnormal there; we can therefore choose as our time
coordinate the proper time of such an observer. As a corollary, we
demand that constant-time slices be flat. We then obtain the condition
\beq
{1 \over 1 - g(r)} - (1-g(r))(f'(r))^2 = 1 \; .	\label{fdiffeqn}
\eeq
There is no need to integrate this; from $dt_s = dt + f'(r)dr$, 
we can read off the Painlev\'e line element:
\beq
ds^2 = - \lf 1 - {2 M \over r} \rt dt^2 + 2 \sqrt{2 M \over r}
dt~ dr + dr^2 + r^2 d \Omega_2^2 \; .
\eeq
Similar coordinate
systems have been found for de Sitter space\cite{tunneldS} and for
black holes in anti-de Sitter space\cite{ads}. The
Painlev\'e metric has a number of attractive features. First, and
crucially, none of
the components of either the metric or the inverse metric diverge at
the horizon. Second, by construction, constant-time slices are just
flat Euclidean space. Third, the generator of $t$ is a Killing
vector. ``Time'' becomes spacelike across the horizon, but is
nevertheless Killing. Finally, an observer at infinity does not make
any distinction between these coordinates and static coordinates;
the function $f$ that distinguishes the two time coordinates
vanishes there.
These coordinates cover the inside and outside of the black hole, or
half the maximally extended space. 

The radial null geodesics in these coordinates obey 
\beq
{dr \over dt} = \pm 1 - \sqrt{2M \over r} \; ,	\label{null}
\eeq
where the plus (minus) sign corresponds to rays that go towards (away from)
infinity. When the particle is inside the black hole, both
ingoing and outgoing trajectories correspond to decreasing $r$, and the
particle cannot (classically) cross the horizon. For massive particles
with worldline tangent $U^a$ we find, using $U_t = -1$, that
\beq
U^t = 1 \To \tau = t + c \; ,
\eeq
and we see that the Painlev\'e time coordinate is precisely the proper
time, $\tau$, for a radially free-falling observer.

These equations are modified when the particle's self-gravitation is
taken into account \cite{kw}. Consider a particle in the s-wave i.e. a shell. 
If the shell has energy $E$, then the geometry inside and outside the
shell are both Schwarzschild spacetimes, but with different mass parameters. 
One can now ask which geometry determines the motion of the
self-gravitating shell. 
It turns out that, when the total energy is held fixed, it is the interior
$E$-dependent metric that determines the motion. That is, we should
replace $M$ with $M-E$ in the geodesic equation.

\section{Tunneling Across the Horizon}
The advantage of having a coordinate system that is well-behaved
at the horizon is that one can study across-horizon physics.
Here we will consider the tunneling of massless shells. The
purpose of truncating to the s-wave is that it is then possible to integrate
out gravity. For spherical gravity, Birkhoff's theorem states that the
only effect on the geometry that the presence of a shell
has, is to provide a junction condition for matching the total mass
inside and outside the shell. In other words, the outgoing shell obeys
Eq. (\ref{null}) with the plus sign, and with $M$ replaced by $M-E$ to
account for self-gravitation.

Now, because of the infinite blueshift near the horizon, the
characteristic wavelength of any wavepacket is always arbitrarily
small there, so that the geometrical optics limit becomes an
especially reliable approximation. 
The geometrical optics limit allows us to obtain rigorous results directly
in the language of particles,\cite{tunnel,nobog} rather than having to
use the second-quantized Bogolubov method.
In the semi-classical limit, we can
apply the WKB formula. This relates the tunneling amplitude to the
imaginary part of the particle action at stationary phase. 
The emission rate, $\Gamma$, is the square of the tunneling amplitude: 
\beq
\Gamma \sim \exp (- 2 ~{\rm Im}~ I) \approx \exp (-\beta E) \; .
\label{gamma}
\eeq
On the right-hand side, we have equated the emission probability to
the Boltzmann factor for a particle of energy $E$. To the extent
that the exponent depends linearly on the energy, the thermal
approximation is justified; we can then identify the inverse
temperature as the coefficient $\beta$.

To calculate the action, first observe that we can formally write it as 
\beq
{\rm Im}~ I = {\rm Im}~ \int_{r_i}^{r_f} p_r ~dr = {\rm Im}~ \int_{r_i}^{r_f}
\int_0^{p_r} dp'_r ~dr \; ,
\eeq
where $p_r$ is the radial momentum. We expect the initial radius,
$r_i$,  to correspond roughly to the site of pair-creation, which
should be slightly inside the horizon, $r_i \approx 2M$. 
We expect the final radius, $r_f$, to be 
slightly outside
the final position of the horizon, else the particle would not be able
to propagate classically to infinity. So $r_f \approx 2(M-E)$.
Because the horizon shrinks, $r_f$ is actually less than $r_i$.
Note how self-gravitation is essential to the tunneling
picture. Without self-gravitation, particles created just inside the
horizon would only have to tunnel just across -- an infinitesimal
separation -- so there wouldn't be any barrier.
But back-reaction results in a shift of the horizon
radius; the finite separation between the initial and final radius is
the classically-forbidden region, the barrier. 

We now eliminate the momentum in favor of energy by using Hamilton's equation
\beq
\left . {d H \over dp} \right |_r = {\pl H \over \pl p} = {dr \over
dt} \; ,
\eeq
where the Hamiltonian, $H$, is the generator of Painlev\'e time. Hence
within the integral over $r$, one can trade $dp$ for $dH$. 
The imaginary part of the action is then 
\beq 
{\rm Im}~I = {\rm Im}~ \int_{r_i}^{r_f} \int_0^H {dH' \over {dr \over
    dt}} dr = -{\rm  Im}~ \int_{r_i}^{r_f} \int_0^E {dr ~ dE' \over 1
  - \sqrt{2(M-E') \over r} } \; , 
\eeq 
where the Hamiltonian, $H$, is just $M-E$, and we have
substituted the self-gravitating radial geodesic for $dr/dt$.
Substituting $u = \sqrt{r}$, and using the
Feynman prescription to displace the energy from $E'$ to $E' - i
\varepsilon$, we have
\beq 
{\rm Im}~I = - {\rm Im}~ \int_{u_i}^{u_f} \int_0^E {2 u^2 du \over u -
  \sqrt{2(M-E' + i \varepsilon)}} ~ dE' \; .
\eeq
and we see that there is a pole in the upper-half $u$-plane. 
The integral can be
evaluated by deforming the contour around the pole. Note that all
real parts, divergent or not, can be discarded since they only
contribute a phase. For example,
the second member of the pair contributes nothing to the tunneling
rate, since it is always classically allowed
and therefore has real action. Doing the $u$
integral first we find
\beq 
{\rm Im}~ I = +4 \pi \int_0^E {d E' (M-E')} \; ,
\eeq
where we used $u_i > u_f$ to obtain the right sign. The tunneling rate
is therefore
\beq 
\Gamma \sim \exp \lf - 8 \pi M E \lf 1 - {E
  \over 2M } \rt \rt = \exp(\Delta S) \; .  \label{rate}
\eeq
To linear order in $E$, we find that the rate is a Boltzmann
factor $\exp (-\beta E)$ with inverse temperature $\beta = 8 \pi
M$. This is the familiar result. But note that at higher energies the
spectrum cannot be approximated as thermal. 
The precise expression, Eq. (\ref{rate}), 
can be written as the exponent of the
difference in the Bekenstein-Hawking entropy, $\Delta S$, before and
after emission \cite{tunnel,peresko}.

Note also that Eq. (\ref{rate}) is consistent with an underlying
unitary theory. For quantum mechanics tells us that
the rate must be expressible as 
\beq
\Gamma (i \to f) = |M_{fi}|^2 
\cdot \lf \mbox{\rm phase space factor} \rt \; ,
\eeq
where the first term on the right is the square of the amplitude 
for the process. The phase space factor is obtained by summing over
final states and averaging over initial states. But the number of final
states is just the final exponent of the final entropy, while the number of
initial states is the exponent of the initial entropy. Hence
\beq
\Gamma \sim {e^{S_{\rm final}} \over e^{S_{\rm initial}}} = \exp
(\Delta S) \; ,
\eeq
in agreement with our result. This suggests that the formula we have
is actually exact, up to a prefactor.

We have found that energy conservation not only supplies the barrier
through which the particle tunnels but also, as anticipated, causes
the spectrum to deviate from exact thermality at higher
energies. However, the form of the correction is not sufficient by
itself to relay information. Consider the emission of two particles
$E_1$ and $E_2$, and the emission of one particle with
their combined energies, $E_1 + E_2$. We find that
\beq
\ln \lf \Gamma_{E_1} \Gamma_{E_2} \rt = -8 \pi \left [E_1 \lf M -
  {E_1 \over 2}   \rt + E_2 \lf M - E_1 - {E_2 \over 2} \rt \right ] =
\ln \Gamma_{E_{1+2}} \; ,
\eeq
so there is no correlation, at least at late-times. It would be very
interesting to see if there are any short-time correlations. In
particular, when a particle is emitted there is a relaxation time for
the black hole to equilibrate. If another particle is emitted during
this time, there might be a correlation that falls off as a function
of the time between the two emissions.

\begin{flushleft}
{\sc Acknowledgments}
\end{flushleft}
\noindent
I would like to thank Dan Kabat for discussions. The author is
supported by DOE grant DF-FCO2-94ER40818.


\begin{thebibliography}{99}

\bibitem{hawking}
 S. W. Hawking, ``Particle Creation by Black Holes,''
 Commun. Math. Phys. {\bf 43} (1975) 199.

\bibitem{infoloss}
S. W. Hawking, ``Breakdown of Predictability in Gravitational
Collapse,'' Phys. Rev. {\bf D14} (1976) 2460.

\bibitem{callanmalda}
C. G. Callan, Jr. and J. M. Maldacena, ``D-Brane Approach to Black
Hole Quantum Mechanics,'' Nucl. Phys. {\bf B472} (1996) 591; {\tt
  hep-th/9602043}.

\bibitem{kvv}
K. Schoutens, E. Verlinde, and H. Verlinde, ``Black Hole Evaporation
and Quantum Gravity,'' {\tt hep-th/9401081};
S. B. Giddings and M. Lippert, ``The Information Paradox and the
Locality Bound,'' {\tt hep-th/0402073}.

\bibitem{tunnel}
M. K. Parikh and F. Wilczek, ``Hawking Radiation as Tunneling,'' 
Phys. Rev. Lett. {\bf 85} (2000) 5042; {\tt hep-th/9907001}.

\bibitem{membrane}
M. K. Parikh and F. Wilczek, ``An Action for Black Hole Membranes'', Phys.
Rev. {\bf D58} (1998) 064011; {\tt gr-qc/9712077};
M. K. Parikh, ``Membrane Horizons: The Black Hole's New Clothes,''
Princeton University Ph.D. thesis; {\tt hep-th/9907002}.

\bibitem{tunneldS}
M. K. Parikh, ``New Coordinates for de Sitter Space and de Sitter
Radiation,'' Phys. Lett. {\bf B546} (2002) 189; {\tt hep-th/0204107};
A. J. M. Medved, ``Radiation via Tunneling from a de Sitter
Cosmological Horizon,'' {\tt hep-th/0207247}.

\bibitem{ads}
S. Hemming and E. Keski-Vakkuri, ``Hawking Radiation from AdS Black
Holes,'' Phys. Rev. {\bf D64} (2001) 044006; {\tt gr-qc/0005115};
E. C. Vagenas, ``Semiclassical Corrections to the Bekenstein-Hawking Entropy
of the BTZ Black Hole via Self-Gravitation,'' Phys. Lett. {\bf B533}
(2002) 302; {\tt hep-th/0109108}.

\bibitem{kw}
P. Kraus and F. Wilczek, ``Self-Interaction Correction to Black Hole
Radiance,'' Nucl. Phys. {\bf B433} (1995) 403; {\tt gr-qc/9408003}.

\bibitem{nobog}
A. J. Hamilton, D. Kabat, and M. K. Parikh, ``Cosmological Particle
Production without Bogolubov Coefficients,'' {\tt hep-th/0311180}.

\bibitem{peresko}
P. Kraus and E. Keski-Vakkuri, ``Microcanonical D-branes and Back
Reaction,'' Nucl. Phys. {\bf B491} (1997) 249; {\tt hep-th/9610045}.

\end{thebibliography}
\end{document}